# Who "Controls" Where Work Shall be Done? State-of-Practice in Post-Pandemic Remote Work Regulation

Darja Smite[1,2], Nils Brede Moe[2,1], Maria Teresa Baldassarre[3], Fabio Calefato[3], Guilherme Horta Travassos[4], Marcin Floryan[5], Marcos Kalinowski[6], Daniel Mendez[1,7], Graziela Basilio Pereira[8], Margaret-Anne Storey[9], Rafael Prikladnicki[8]

[1] Blekinge Institute of Technology (Sweden), [2] SINTEF (Norway), [3] University of Bari (Italy), [4] Federal University of Rio de Janeiro (Brazil), [5] Independent (Sweden), [6] Pontifical Catholic University of Rio de Janeiro (Brazil), [7]fortiss (Germany), [8] Pontifical Catholic University of Rio Grande do Sul (Brazil), [9] University of Victoria (Canada)

ABSTRACT

The COVID-19 pandemic has permanently altered workplace structures, making remote work a widespread practice. While many employees advocate for flexibility, many employers reconsider their attitude toward remote work and opt for structured return-to-office mandates. Media headlines repeatedly emphasize that the corporate world is returning to full-time office work. This study examines how companies employing software engineers and supporting roles regulate work location, whether corporate policies have evolved in the last five years, and, if so, how, and why. We collected data on remote work regulation from corporate HR and/or management representatives from 68 corporate entities that vary in size, location, and orientation towards remote or office work. Our findings reveal that although many companies prioritize office-centred working (50%), most companies in our sample permit hybrid working to varying degrees (85%). Remote work regulation does not reveal any particular new "best practice" as policies differ greatly, but the single most popular arrangement was the three in-office days per week. More than half of the companies (51%) encourage or mandate office days, and more than quarter (28%) have changed regulations, gradually increasing the mandatory office presence or implementing differentiated conditions. Although no companies have increased flexibility, only four companies are returning to full-time office work. Our key recommendation for office-oriented companies is to consider a trust-based alternative to strict office presence mandates, while for companies oriented toward remote working, we warn about the points of no (or hard) return. Finally, the current state of policies is clearly not final, as companies continue to experiment and adjust their work regulation.

## 1. Introduction

The COVID-19 pandemic has triggered irreversible changes in the workplace, with remote work becoming an integral part of modern society (Smite et al., 2023; Allen et al., 2024). However, notable tensions exist between employees seeking flexibility and employers pushing for structured return-to-office mandates (Ding and Ma, 2023; Eng et al., 2024). It can be illustrated in an overview of the publicly announced changes in the big tech companies (See Fig. 1[1]). A similar trend is reported in the recent Developer Survey by Stack Overflow[2], which captures a continuous growth of in-person working in the last three years from 15% of developers in 2022 to 16% in 2023 and with an even larger jump to 20% in 2024. The recent trends can be found in the employee forum Blind[3] that became the go-to chronologer of the changes in work regulation and in the hybrid working platform FlexIndex[4], which became the go-to catalog of work arrangements with records from over 10.000 companies worldwide. The debate of whether remote work is a right or a privilege (Smite et al., 2023) is ongoing, attracting attention, and has even led to legal disputes (Delbosc and Kent, 2024).

At the core of the conflict between employees and employers lies a fundamental shift in beliefs about where work should be done and how companies should regulate it. Many employees view remote work as a necessary adaptation that enhances productivity and work-life balance. In contrast, employers, particularly top management, often argue that in-person presence is crucial for maintaining employee engagement and organizational performance (Delbosc and Kent, 2024; Eng et al., 2024; Eng et al., 2025).

Return-to-office policies are blamed as means of regaining control over employees (Ding and Ma, 2023). As such, they have resulted in significant pushback (Barrero et al., 2021; Delbosc and Kent, 2024; Ding and Ma, 2024). Barrero et al. (2021a) report that 36% of employees would consider finding a new job, and 6% would quit if forced to return to the office full-time. Further, empirical studies show a significant decline in employee job satisfaction after US firms announced a return-to-office mandate (Ding and Ma, 2024) and an abnormally high employee turnover (Ding et al., 2024). These findings suggest that inflexible workplace policies could lead to talent loss, lower morale, and difficulty in attracting new talent, ultimately impacting long-term business performance.

[1] Based on the records by BuildRemote (https://buildremote.co)
[2] 2024 Developer Survey by Stack Overflow (https://survey.stackoverflow.co/2024/work#employment)
[3] A community forum for professionals across companies and industries that facilitates communication (www.teamblind.com)
[4] A commercial insight platform on flexible work covering tens of thousands of companies worldwide (www.flexindex.com)

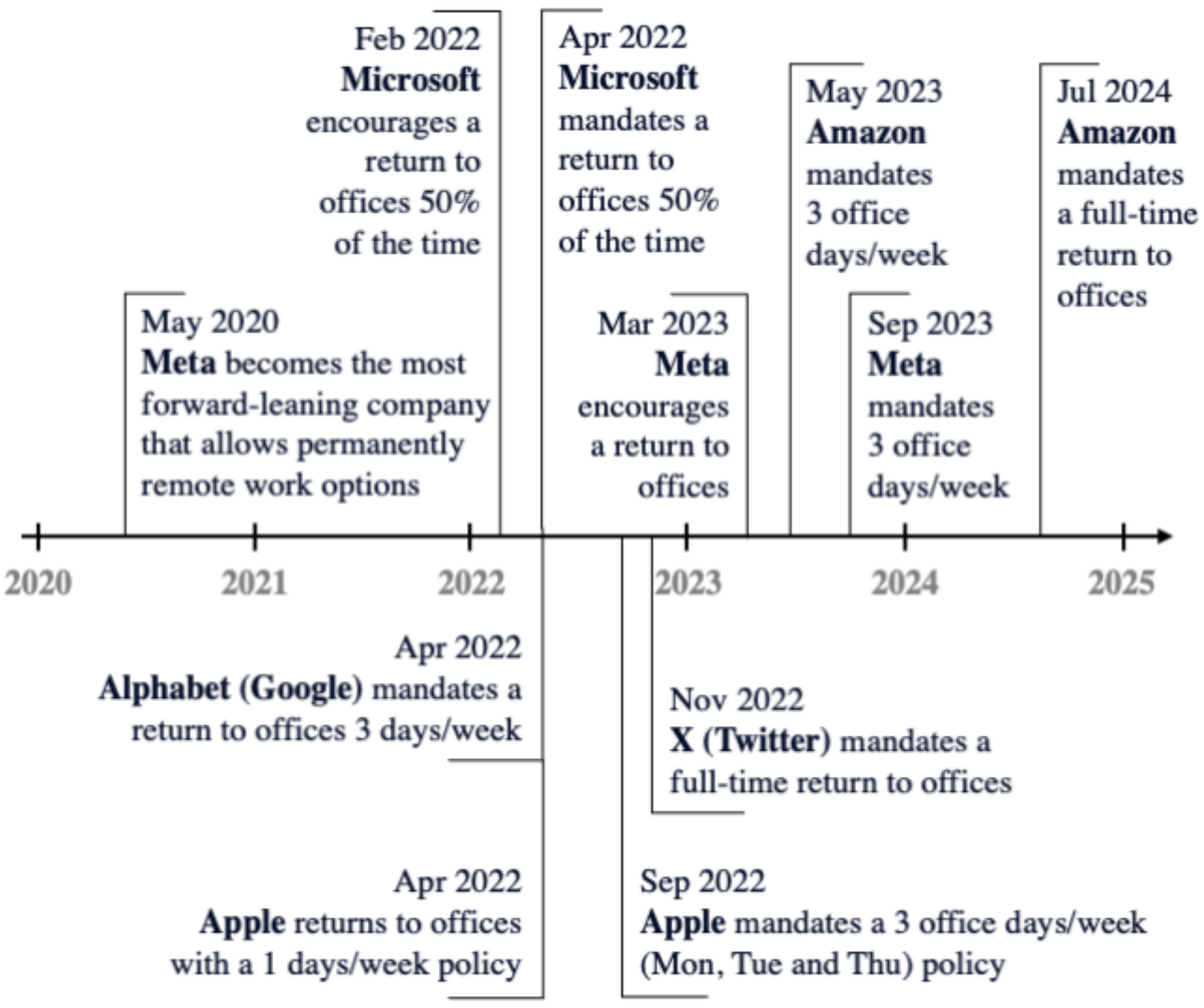

**Fig. 1:** Overview of the publicly announced changes in remote work regulation in the big tech companies

The latest insights into the post-pandemic workplaces emphasize the inherent struggle that, on one hand (employees' perspective), is seen as the fight for freedom from managerial control, and on the other hand (employers' perspective), is seen as the fight for collaboration, engagement and organizational performance. The future seems to depend on how organizations will adapt their policies to balance employee and organizational needs.

In this paper, we explore these workplace trends in companies developing software-intensive products and services. The following question drives our work:

*How is work location regulated, and have companies changed their internal approaches to regulating remote and onsite work in recent years?*

Our study reveals the latest trends in remote work regulation. We found a variety of nuanced strategies rather than convergence on a particular best practice. We also report the evolution of work regulation and how some companies shifted their return-to-office (RTO) mandates to intensified office presence, while others implemented more subtle adjustments. Finally, we report that only a small minority of companies reverted to full-time office work.

The rest of the paper is organized as follows. In Section 2, we outline related studies on emerging work regulation. Section 3 details our empirical cases, data collection, and analysis. Section 4 presents our results, followed by a discussion in Section 5, and conclusion in Section 6.

## 2. Background – Emerging work regulation

In our previous study of emerging work regulation (Smite et al., 2023), conducted in early 2022, we analyzed 26 post-pandemic work policies from 17 companies and their sites, covering 12 countries. Based on the initial work policies, we concluded that there is a great variation in how much flexibility companies are willing to yield to their employees, with half of the companies restricting remote working and half of the companies having no centralized regulation of remote work or office presence. Our work was heavily biased towards Scandinavian companies and reflected the state of the practice in the early post-pandemic months.

Challenges in implementing a consistent office-first strategy in three companies have been documented by Smite and Moe (2022). Their results suggest that certain post-pandemic changes in the workplace may discourage employees from working onsite. These include downsized offices, bookable desks, and cancelled free parking. At the same time, companies experiment with different approaches to make the office more attractive and support their office-centric strategies with in-office events, classes, and sports activities, a better canteen, convenient work areas, and improved noise isolation (Smite and Moe, 2022).

RTO mandates have led to a public debate and attracted researchers' attention. Pandita et al. (2024) have studied the psychological effects of RTO mandates and found that mandated office work can cause increased emotional exhaustion and a rising tendency of presenteeism, which is linked with the differences in how onsite vs. remote employees are treated. The authors report such effects as fears of negative outcomes for remote work and perceived lack of trust towards remote workers, among other factors. The lack of trust and willingness to regain control as the motivation for RTO mandates is also reported by Ding and Ma (2023) who studied public RTO announcements in 134 firms. Differences in treatment of remote employees surface among peers too, as found by Tkalich et al. (2024), who reported challenges with maintaining psychological safety in dispersed teams, and increasing feelings of alienation and exclusion (marked as "second-class citizenship") among remote team members.

An extensive analysis of over three million tech and finance workers' employment histories from LinkedIn (Ding et al., 2024) suggests abnormally high employee turnover following RTO mandates, especially among females, seniors, and more skilled employees. The study also reports that companies with RTO mandates have challenges filling their job vacancies, implying that the changes in work regulation significantly damage the corporate attractiveness as an employer.

The effect of RTO mandates has been studied using resumes matched to company data at Microsoft (Van Dijcke et al., 2024). The stratified analysis of the employees' tenure and seniority suggests that the shift to 50% office presence is followed by increased resignations, with higher rates among more senior employees, including managers.

While the perception and influence of RTO mandates have gained a lot of attention, few studies have addressed the employers' perspectives. Eng et al., (2025) interviewed 17 managers from SMEs in Sweden and emphasized the dual effects of flexible policies, and especially the challenges they pose to sustaining innovation, productivity, and organizational cohesion.

In this paper, we build upon previous research by examining corporate remote work regulations, strategies, expectations, and motivations from the perspective of employers across companies of all sizes—from small businesses to large enterprises—worldwide. Our goal is to explore how organizations have adapted to evolving remote work trends five years after the pandemic, now that big tech companies are setting the stage and, in many cases, governments have established legislation for regulating remote work.

## 3. Overview of the study and empirical cases

The results are based on an extensive empirical multi-case study of remote work regulation in companies that develop or maintain software-intensive systems and services. The case selection followed a non-random convenience sampling without restricting conditions regarding the size, location, application domain, or remote work regulation in line with a maximum variation strategy (Flyvberg, 2006). The cases were selected based on availability in the professional network of participating researchers.

The final dataset comprises 68 cases (see Table 1), with a fair distribution in terms of company size, domain and locations (see Fig. 2). Admittedly, the dataset is slightly biased towards very large European companies and those operating as software development consultancies.

Data collection followed a structured protocol. Company representatives were asked to share or show their work policy documents, if those existed. Each researcher was responsible for collecting information based on an interview guide and could conduct an in-person interview, an online interview, or offline data collection via email exchange. During the interview (if conducted in-person or online), researchers took notes to document responses to the questions, while offline data collection led to soliciting answers to questions from the representatives in written form directly.

Data analysis started by categorizing the work regulation according to the characteristics described in Fig. 3 and plotting these characteristics in summarizing graphs (See Fig. 4 - Fig. 7 and Fig. 9). The summaries were then used to identify trends. Further analysis was done in iterations, going back and forth between the summaries, individual case policies, and interview notes to find explanations for the observed trends. We performed a detailed analysis of the motivations for the chosen work regulation (See Fig. 8) by manually coding the responses given by the interviewees and categorizing them into two groups – those in favor of remote work (111 codes) and those in favor of in-office work (58 codes). Similarly, we have summarised the prevalence of fully remote positions and hiring strategies in the studied companies stratified by orientation towards remote wor, office work, or no orientation (See Fig. 10). During the analysis, researchers met to discuss the major findings and decide on the direction of further analysis.

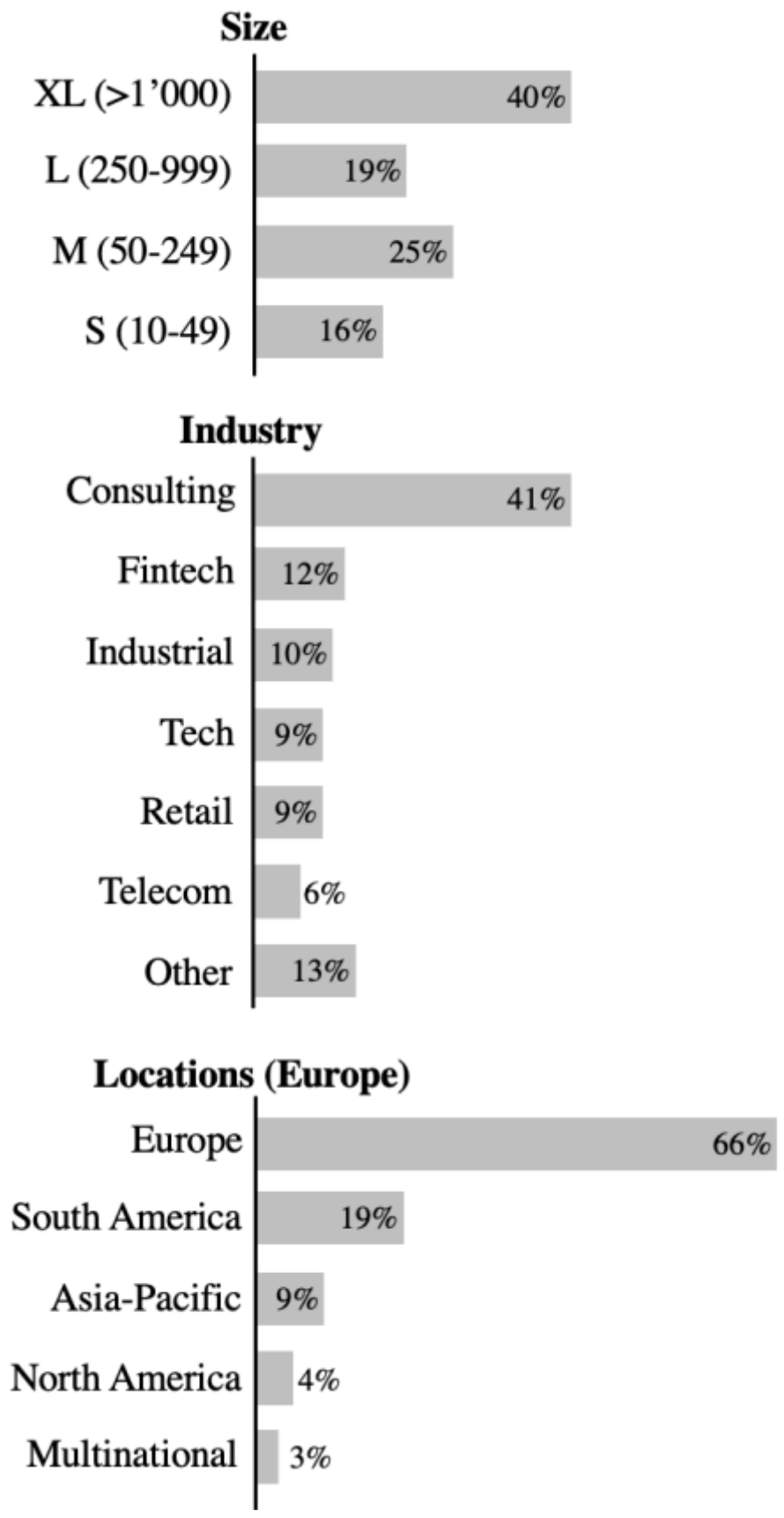

**Fig. 2**: Demographic characteristics of the cases

The process of engaging with companies was structured as follows: Each participating researcher recruited at least two companies from their professional network. They introduced the study's objectives, research design, and data management plan, conducted interviews, addressed clarification requests, validated the interpretation of the findings, secured company approval for the final report, and presented the final results to company representatives.

Noteworthy, the use of convenience sampling may have introduced bias, as the sample may overrepresent certain types of work regulations from very large companies (40% in our sample) and underrepresented others, for example, US-based companies (only three in our sample), reducing the generalizability of findings. Nonetheless, the results provide valuable directional insights into emerging trends and shared motivations.

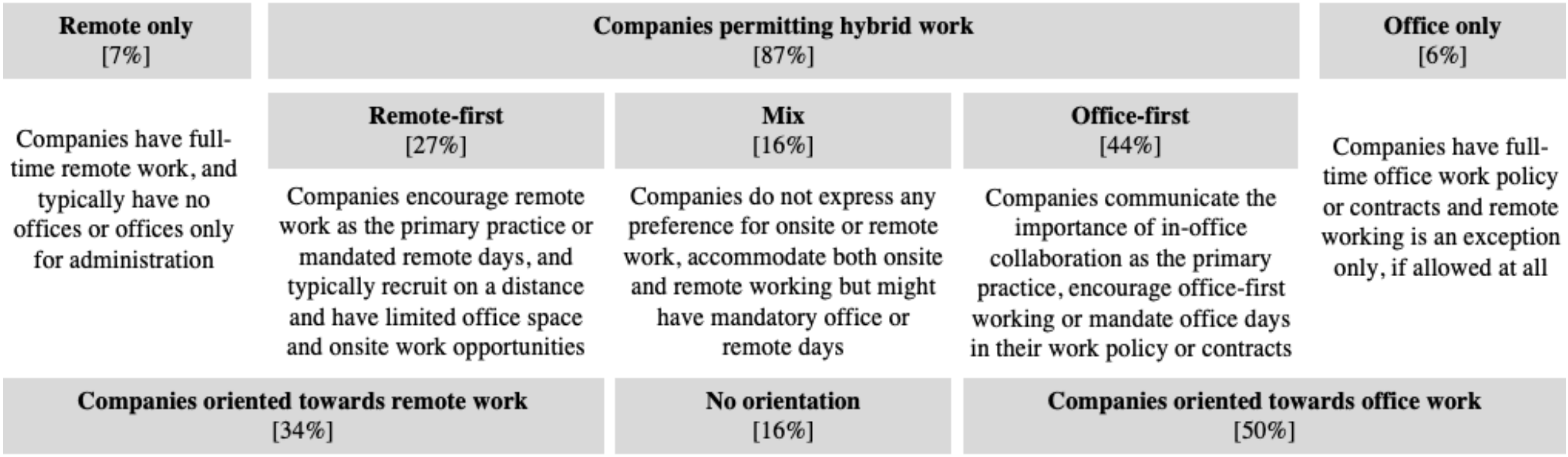

**Fig. 3**: Categorization of work regulation used in the data analysis to summarize the cases and the number of cases per category

**Table 1:** Overview of the cases and data collected (sorted by location, size[5], and industry).

| Size | Locations | C | Company | Domain | Documentation | Form | Interviewee(s) | Date |
|---|---|---|---|---|---|---|---|---|
| XL | APAC | C66 | [redacted] | Consulting | – | Offline | Regional Head of People located in China | 03/2025 |
| XL | Brazil | C2 | Petrobras | Industrial | – | Offline | Head of Digital Transformation | 11/2024 |
| XL | Brazil | C3 | Eletrobras | Industrial | – | In-person | General Manager of Digital Transformation | 11/2024 |
| XL | Brazil | C6 | Stone Co | Fintech | – | In-person | Head of R&D | 11/2024 |
| XL | Brazil | C1 | [redacted] | Consulting | Policy document | Online | Head of People LATAM | 11/2024 |
| XL | Brazil | C61 | [redacted] | Tech | – | Offline | Head of People | 03/2025 |
| XL | Brazil | C63 | [redacted] | Consulting | – | Offline | Talent Business Partner | 03/2025 |
| XL | Brazil | C64 | [redacted] | Consulting | – | Offline | Talent Business Partner | 03/2025 |
| XL | Europe | C68 | [redacted] | Consulting | – | Offline | Head of People | 03/2025 |
| XL | India | C67 | [redacted] | Consulting | – | Offline | Head of People | 03/2025 |
| XL | Italy | C19 | [redacted] | Consulting | | Online | Head of Talent Management & People Development (HR) | 10/2024 |
| XL | Mexico | C29 | [redacted] | Industrial | – | Online | Leader of Training Programs | 11/2024 |
| XL | Multinational | C15 | [redacted] | Content and media | Guidelines, Playbook | Online | Engineering manager | 11/2024 |
| XL | Multinational | C57 | [redacted] | Tech | – | In-person | Head of HR Operations | 03/2025 |
| XL | Norway | C31 | Multiconsult | Industrial | – | Online | Head of Property Management | 11/2024 |
| XL | Norway | C32 | [redacted] | Telecom | Handbook document and FAQ | Online | Principal Software Engineer and HR Manager | 03/2025 |
| XL | Norway | C33 | Deloitte NO | Consulting | Policy document | Online | Director, verified with the HR manager | 03/2025 |
| XL | Norway | C34 | Storebrand | Fintech | Policy document | Online | HR manager | 10/2024 |
| XL | Norway, Sweden | C36 | [redacted] | Retail (online) | – | Online | VP of Transformation | 03/2025 |
| XL | Poland | C39 | VOX Capital Group | Retail (in-house IT) | Policy | Online | CTO and head of HR | 11/2024 |
| XL | Sweden, Estonia,Latvia Lithuania | C43 | [redacted] | Fintech | – | Online | Chief Competence Lead and HR Responsible for Work Experience | 12/2024 |
| XL | Sweden | C44 | [redacted] | Telecom | Policy, Announcement | In-person | Facility manager | 11/2024 |
| XL | Sweden | C45 | Telenor SE | Telecom | – | In-person | Facility manager | 11/2024 |
| XL | Sweden, Netherlands | C46 | [redacted] | Retail (in-house IT) | Handbook document. Internal presentation. | Online | Product owner | 11/2024 |
| XL | Sweden | C47 | [redacted] | Retail (in-house IT) | – | Online | Workspace experience lead | 11/2024 |
| XL | Sweden | C48 | Tietoevry | Consulting | Guidelines document | Offline | Head of HR Operations and HR Director | 02/2025 |
| L | Brazil | C4 | Capemisa | Fintech | Policy document | Online | HR Manager and HR team | 10/2024 |
| L | Brazil | C5 | Maxtrack | Logistics | Policy document | Online | CEO, HR Manager, and HR team | 10/2024 |

[5] Company size is determined using the following scale: XS – micro (less 10 people), S – small (10-49 people), M – medium (50-249 people), L – large (250-999 people), XL – very large (more than 1000 people)

| | | | | | | | | |
|---|---|---|---|---|---|---|---|---|
| L | Brazil | C62 | [redacted] | Tech | – | Offline | Head of People | 03/2025 |
| L | Germany, Poland | C11 | QualityMinds | Consulting | Policy document | Offline | CEO | 02/2025 |
| L | India | C16 | [redacted] | Telecom | Initial policy, updated policy announcement | Offline Online | Engineering Manager (Sweden) | 01/2025 03/2025 |
| L | Italy | C20 | [redacted] | Govtech | Policy document | Online | Head researcher of the local site | 11/2024 |
| L | Italy | C21 | [redacted] | Consulting | | Online | Deputy HR Director | 11/2024 |
| L | Italy | C59 | [redacted] | Fintech | – | Online | HR manager | 03/2025 |
| L | Italy | C55 | [redacted] | Retail (in-house IT) | Policy document | Online | Head of HR | 03/2025 |
| L | Finland | C65 | [redacted] | Retail (online) | – | Online | VP of Transformation | 03/2025 |
| L | Norway | C35 | SB1 Utvikling | Fintech | Policy document | Online | HR manager | 10/2024 |
| L | Norway | C37 | Knowit | Consulting | Policy document | Online | HR manager | 10/2024 |
| L | Norway | C58 | Avinor | Industrial | Policy | In-person | Line manager | 02/2025 |
| L | Norway, India, China, Germany | C17 | DNV | Industrial | – | In-person | Local Site Manager and Line Manager (head of digital development department) | 02/2025 03/2025 |
| M | Brazil | C7 | DB | Consulting | – | Online | One of the founders | 11/2024 |
| M | Denmark, Sweden | C10 | [redacted] | Consulting | Guide document and Playbook document. | Online | CTO | 11/2024 |
| M | Germany | C12 | Keil KTM | Consulting | – | Online | Managing Director | 11/2024 |
| M | Germany, US | C13 | CQSE | Consulting | Policy document | Offline | Founding partner | 01/2025 |
| M | India | C18 | [redacted] | Fintech | Guideline document. Policy document. | Online | Head of product development | 11/2024 |
| M | Italy | C22 | Hevolus | R&D | Policy document | In person | Head of Software Engineering | 11/2024 |
| M | Italy | C23 | [redacted] | Consulting | | Online | Local Site Manager | 11/2024 |
| M | Italy | C24 | Klopotek | Content and media | Policy announcement | Online | Site manager | 10/2024 03/2025 |
| M | Norway | C38 | Kantega | Consulting | Policy document | Online | Two representatives from the leadership team | 10/2024 |
| M | Norway | C60 | Trondheim Digital | Govtech | Policy document | Offline & online | CTO | 03/2025 |
| M | Poland | C40 | [redacted] | Consulting | Policy | Online | Head of People and Culture | 11/2024 |
| M | Poland | C41 | [redacted] | Consulting | Policy | Online | Chief HR Officer | 11/2024 |
| M | Sweden | C49 | [redacted] | Fintech | Guidelines and Policy | Online | Head of engineering | 11/2024 |
| M | Sweden | C50 | [redacted] | Industrial | – | Offline | Head of Operations, and Head of a daughter company | 11/2024 |
| M | Sweden | C51 | Synteda | Consulting | – | Online | CEO | 01/2025 |
| M | UK | C54 | HST | Content and media | Manifesto document | Offline | HR director | 12/2024 |
| M | US | C56 | [redacted] | Tech | – | Offline | CEO | 03/2025 |
| S | Brazil | C8 | Lemobs | Govtech | Policy document | Online & offline | CEO and HR Manager | 10/2024 |
| S | Brazil | C9 | Webdraw | Tech | – | Offline | CEO and co-founder | 02/2025 |
| S | Germany | C14 | Improuv | Consulting | – | Online | Managing Partner | 11/2024 |
| S | Italy | C25 | Apuliasoft | Consulting | Handbook document | In-person | Chief HR & Happiness Officer | 10/2024 |
| S | Italy | C26 | Ai2 | Consulting | Drafted guidelines | In-person | CEO | 10/2024 |
| S | Italy | C27 | [redacted] | Consulting | – | Online | CEO / Owner of the company | 11/2024 |
| S | Italy | C28 | SER&Practices | R&D | Policy document | In person | Production Manager | 112024 |
| S | New Zealand | C30 | Multitudes | Tech | Policy document | Offline | Chief of Staff | 03/2025 |
| S | Poland | C42 | Lunar Logic | Consulting | Working agreements | Online | CEO | 11/2024 |
| S | Sweden | C52 | Malvacom | Consulting | – | In-person | Manager | 12/2024 |
| S | Sweden | C53 | factor10 | Consulting | – | Online | CTO | 11/2024 |

# 4. Results

## 4.1. Current state of remote work regulation

The analysis of the 68 corporate cases studied suggests that remote work regulation is highly diverse with companies divided between those who have central regulation of office and/or remote days (mandatory or recommended) and those who have decentralized the remote work regulation to teams, departments, projects or individuals (employee choice) (see y-axis in Fig. 4). Although central regulation represents a larger group in our sample, with 51% of companies regulating office days (min one and max all five workdays) and 9% having no office work at all, companies with decentralized regulation letting departments or groups decide (19%) and those letting employees decide (21%) are also considerably many (39% in total).

An interesting and perhaps unexpected finding is that the choice of regulation and the number of office or remote days did not necessarily predict the overall management vision or attitude towards the preferred work location. For example, we found companies that expressed the desire or ambition for employees to return to the office, but their work policies had very few mandatory office days or only recommended office presence (to be discussed in Section 4.2). Similarly, we found companies that positioned themselves as remote-first but recommended or required employees to appear in the office 1-2 days per week, often to justify having an office (to be discussed in Section 4.3). Based on this finding, we have arranged the companies according to the general expectation for where the work shall be done (corporate strategy, desire, or ambition), including those oriented towards office work, remote work, and companies with no expressed preference or orientation in the middle (see x-axis in Fig. 4).

When looking at the positioning of the companies in the work regulation space (Fig. 4), one important finding is that despite the heated public debate and recent increase in demanded office presence in the US five years after the pandemic (Fig. 1), only four companies in our sample have returned to the five-day work in the office. Companies that orient towards office work in our sample are a large group, half (50%), with the largest category in terms of regulated office days being three office days/week (24%), according to the current state of remote work regulation.

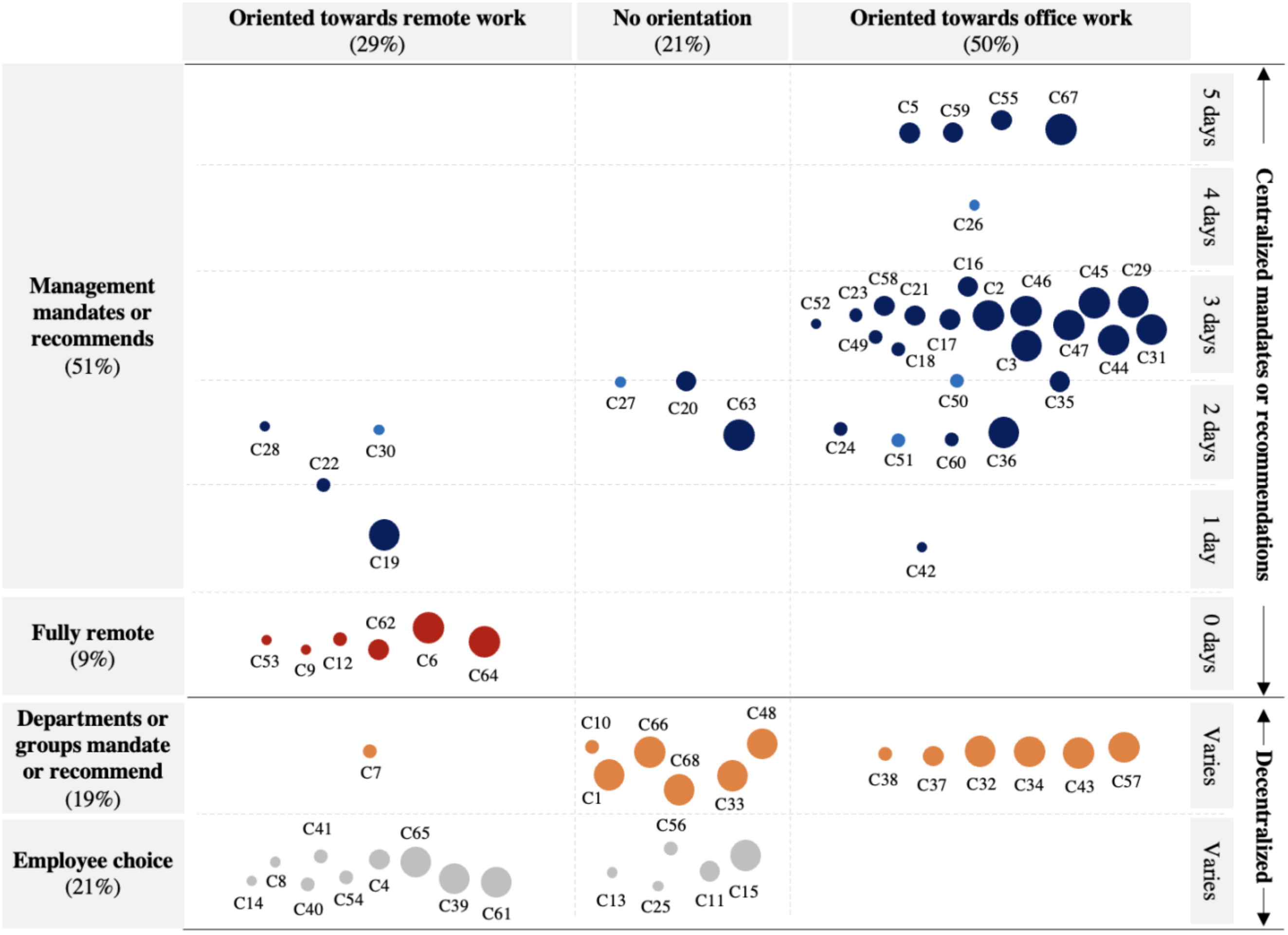

**Fig. 4:** Overview of the required centralized vs decentralized regulation with the corresponding number of centrally mandated (mandatory) or recommended office days on the Y-axis, and the general expectation for where the work shall be done (orientation towards remote work, office work, or no orientation) on the X-axis. The colour indicates the work regulation category: RTO mandates, Recommendation-based, Decentralized to departments, projects or teams, Employee-choice, and Fully remote.

Companies are visualized based on size: Small Medium Large Extra large.

In the following sections, we detail work regulation in companies oriented towards office work, companies oriented towards remote work, and companies with no orientation, revealing nuanced formulations of remote and in-office work regulation and corporate recruitment strategies.

## 4.2. Companies oriented towards office work

In Fig. 5, we summarize the requirements for office presence in companies oriented toward office work.

Office-first companies form the largest category in our sample, representing 50% of the companies studied. Within this group, we identified several distinct regulatory approaches that express expected office presence, ranging from strictly mandated office days to more flexible recommendations (note the different shades of blue that distinguish these two types of regulation in the figure).

The most common pattern among office-first companies was the requirement of three in-office days per week (around 1/4 of companies in our sample). This approach reflects the general trends in the big tech companies like Apple, Microsoft, and Meta, requiring office attendance of 50-60% (see Fig. 1) and combines flexibility with frequent in-person collaboration.

Three companies (Italian C21, Mexican C29, and Indian C16) with three-day office requirements further regulated which specific days employees must be present, indicating a stronger emphasis on synchronized team presence and balanced office space utilization. An interesting case is Indian C16, which requires precisely three in-office days and two remote days per week, and specific days are selected for selected teams. This structured approach was implemented due to insufficient office space.

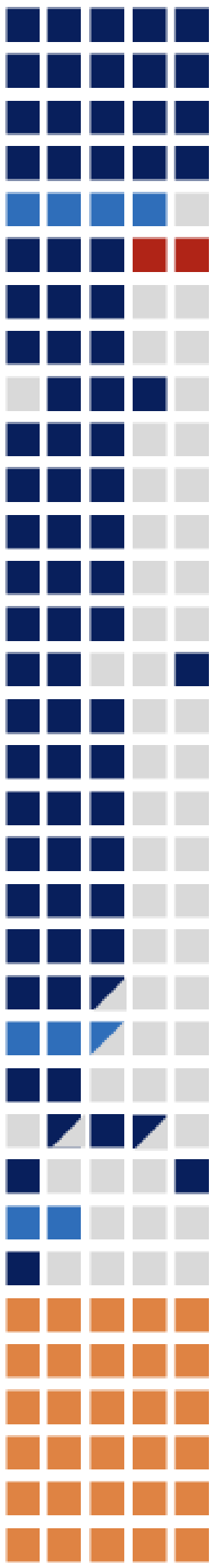

| Onsite/remote days | Case | Size | Location | Required office presence | Fully remote positions | Recruitment |
|---|---|---|---|---|---|---|
| | C67 | XL | India | Full time onsite work | Yes | Within the country |
| | C5 | L | Brazil | Fully time onsite work with exceptional remote positions | Yes | Within commute distance |
| | C59 | L | Italy | Full time onsite work | No | Within commute distance |
| | C69 | L | Italy | Full time onsite work | No | Within commute distance |
| | C26 | S | Italy | 4 days/week of onsite work are recommended | No | Within the country |
| | C16 | XL | India | 3 onsite days and 2 remote days/week required, leaders schedule onsite and remote days | No | Within commute distance |
| | C2 | XL | Brazil | 3 days/week onsite. Immediate managers are mandated to request more office presence | No | Within the country |
| | C3 | XL | Brazil | 3 days/week onsite. Leadership work fully onsite | No | Within commute distance |
| | C29 | XL | Mexico | Tuesdays, Wednesdays and Thursdays are onsite days | Yes | Within the country |
| | C46 | XL | Sweden | Majority of the time of working onsite | No | Within the country |
| | C45 | XL | Sweden | 3 days/week of onsite work required, leaders mandated to decide which are office days | As an exception | Within the country |
| | C47 | XL | Sweden | Majority of the time onsite | As an exception | Within commute distance |
| | C44 | XL | Sweden | 3 days/week of onsite work required | As an exception | Within commute distance |
| | C31 | XL | Norway | At least 3 days/week in the office | As an exception | Within the country |
| | C21 | L | Italy | 3 days/week of onsite work, one of the office days must be either Monday or Friday | As an exception | Within the country |
| | C17 | L | Nor., China, India | 60% onsite work required (3 days / week) | As an exception | Within the country |
| | C58 | L | Norway | 3 days/week onsite | No | Within the country |
| | C23 | M | Italy | 3 days/week onsite | No | Within the country |
| | C49 | M | Sweden | 60% onsite work required (3 days / week) | As an exception | Within commute distance |
| | C18 | M | India | 60% onsite work required (3 days / week) | As an exception | Within commute distance |
| | C52 | S | Sweden | No less than 3 days / week of onsite work | As an exception | Within the country |
| | C35 | L | Norway | At least 50% of time onsite, including 1 common day in a team | As an exception | Within the country |
| | C50 | M | Sweden | 2-3 days/week recommended to work onsite. Managers are mandated to decide | As an exception | Within commute distance |
| | C36 | XL | Norway, Sweden | 2 office days/week onsite | As an exception | Within commute distance |
| | C24 | M | Italy | 2 office days/week onsite. Wed is a mandatory office day, second day – Tue or Thu | As an exception | Within the country |
| | C60 | M | Norway | 2 office days/week onsite – Monday or Friday | As an exception | Within commute distance |
| | C51 | M | Sweden | 2 office days/week recommended | Yes | Within the country |
| | C42 | S | Poland | Office presence on Mondays is expected | No | Within commute distance |
| | C57 | XL | Multinational | Departments decide. Some are 3 days/week onsite, but software developers are flexible | Yes | Within the country |
| | C34 | XL | Norway | Flexible but not fully remote. Teams and departments are mandated to decide | No | Within commute distance |
| | C32 | XL | Norway | No central requirement, departments and teams mandated to decide | No | Within commute distance |
| | C43 | XL | Sweden, the Baltics | Managers are mandated to decide. Some teams/departments have 2 or 3 office days/week | No | Within the country |
| | C37 | L | Norway | Not fully remote. Customers decide. Mostly onsite for those not on a project | As an exception | Within the country |
| | C38 | M | Norway | Not fully remote. Team and customers decide. Mostly onsite for those not on a project | As an exception | Within commute distance |

**Fig. 5:** Work regulation in office-first companies.
Each row represents a company's in-office work requirements across the workdays of a typical week (Mon–Fri), illustrated through color-coded blocks: Mandated office days, Recommended office days, Flexible days (employee-choice), Days with decentralized office presence, subject to department, project or team decisions, and expected remote days. In some cases, specific weekdays are enforced, while in others, only the number of required in-office days per week is prescribed.

Concrete workdays of co-location are also mandated in other companies with lower office presence requirements (Italian 24, Polish LunarLogic (C42), and Norwegian Trondheim Digital (C60)). In this context, the approach chosen by Norwegian SB1 Utvikling (C35) is particularly noteworthy. It requires one common day each team selects for collaboration, balancing individual flexibility with team cohesion needs.

Office presence in the Norwegian Multiconsult (C31) is more flexible than in many other companies. Employees can choose which office to attend in a range of offices throughout the country that were set up to minimize the commute.

Only four companies in our entire sample (Brazilian Maxtrack (C5), Italian C59 and C55, and Indian C67) mandated five office days per week, suggesting that very high office presence requirements remain uncommon even in office-first companies.

Not all office-first companies rely on mandatory presence requirements. Three companies in this group (Swedish C50 and Synteda (C51), and Italian Ai2 (C26)) announce the desire to see their employees working in the office, as an encouragement rather than a strict mandate of office attendance.

Six of the office-first companies delegate decisions about office presence to teams, departments, or projects. The decentralized approach is chosen by four large companies that do not believe in one-size-fits-all solutions (C57, C32, Storebrand (C34), and C43), and two smaller consultancy companies (Knowit (C37) and Kantega (C38)), where customer relationships influence work arrangements. In these consultancy companies, employees without active customer projects or between assignments are generally expected to be present in the office, creating a differentiated regulation where office presence varies based on project status. Notably, all six companies are headquartered in Scandinavia.

An important finding was that local labor laws in countries such as Brazil, Italy, Norway, and Sweden influenced companies' choice for having office presence of over 50%. In Italy, companies allowing remote work for more than 50% of the time have additional administrative burdens and data security requirements. In Sweden, in-office presence is motivated by the desire to avoid legally required control of the suitability of individual home offices.

Overall, office-first companies demonstrate considerable variability in their specific approaches to regulating office presence. It reveals that even within an office orientation, companies adapt their policies to accommodate practical constraints and employee preferences while maintaining their core office-centric philosophy or corporate culture.

## 4.3. Companies oriented towards remote work

The group of companies oriented towards remote work (see Fig. 6) comprises two key categories of companies united by the general orientation towards remote work. The typical characteristics of the companies in this group include decreased office space and the availability or prevalence of fully remote positions.

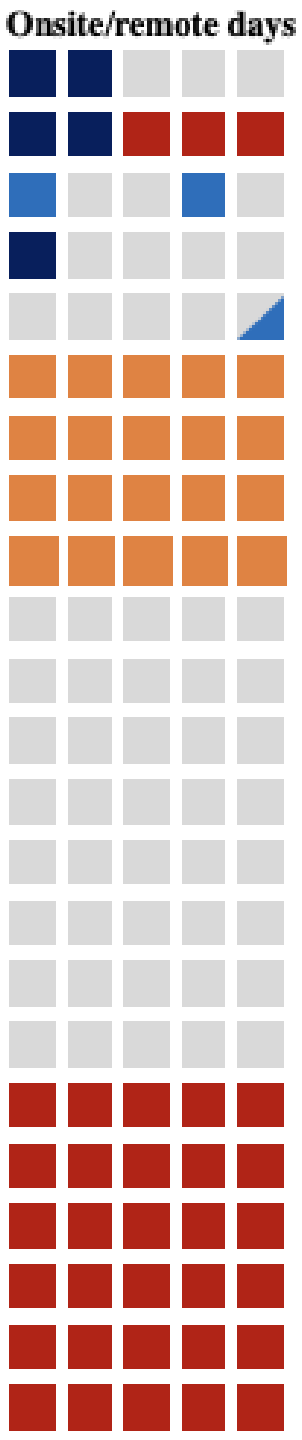

| Case | Size | Location | Required office presence | Fully remote positions | Recruitment |
|---|---|---|---|---|---|
| C22 | M | Italy | 1-2 days/week of onsite work required | Yes | Within the country |
| C28 | S | Italy | 2 fixed days/week in the office and 3 days/week from home | As an exception | Within the country |
| C30 | S | New Zealand | Employees are encouraged to work on Mondays and Thursdays from the office | Yes | Within the country |
| C19 | XL | Italy | 1 onsite day/week required. Office days can be accumulated and redistributed over month | As an exception | Within the country |
| C14 | S | Germany | Every second Friday onsite, presence is encouraged | No | Within commute distance |
| C1 | XL | Brazil | Employee choice unless customers require office presence | Yes | Internationally |
| C66 | XL | APAC | Employee choice unless customers require office presence | Yes | Within the country |
| C68 | XL | Europe | Employee choice unless customers require office presence | Yes | Within the country |
| C7 | M | Brazil | Employee choice, but people living nearby are recommended to work 2 days/week onsite | Yes | Internationally |
| C39 | XL | Poland | Employee choice | Yes | Internationally |
| C61 | XL | Brazil | Employee choice | Yes | Within country |
| C4 | L | Brazil | Teleworking, onsite work on demand, upon authorization | Yes | Within the country |
| C65 | L | Finland | Employee choice | Yes | Within the country |
| C54 | M | The UK | Only planned occassions | Yes | Within the country |
| C40 | M | Poland | Employee choice | Yes | Within the country |
| C41 | M | Poland | Employee choice | Yes | Within the country |
| C8 | S | Brazil | On demand | Yes | Within the country |
| C6 | XL | Brazil | Fully remote, office for administration primarily, available for onsite work, if desired | Yes | Within selected countries |
| C64 | XL | Brazil | Fully remote, office for administration only | Yes | Internationally |
| C62 | L | Brazil | Fully remote, office for administration primarily, available upon authorization | Yes | Within the country |
| C12 | M | Germany | Fully remote, office for administration only | Yes | Within selected countries |
| C9 | S | Brazil | Fully remote, office for administration primarily, available for onsite work, if desired | Yes | Internationally |
| C53 | S | Sweden | Fully remote, no office | Yes | Within the country |

**Fig. 6:** Work regulation in companies oriented towards remote work.
Each row represents a company's in-office work requirements across the workdays of a typical week (Mon–Fri), illustrated through color-coded blocks: Mandated office days, Recommended office days, Flexible days (employee-choice), Days with decentralized office presence, subject to department, project or team decisions, and expected remote days. In some cases, specific weekdays are enforced, while in others, only the number of required in-office days per week is prescribed.

The first category comprises companies that promote remote working, permit fully remote positions, and maintain offices, often of limited capacity, or coworking spaces. Few of the companies in this group require office presence (see the top five companies in Fig. 6). Two of these companies (Italian Hevolus (C22) and SER&Practices (C28)) require two office days/week, one company (Multitudes (C30) in New Zealand) encourages two office days/week, one company (Italian C19) mandates one office day/week, and one company (German Improuv (C14)) encourages onsite work every second Friday. Yet, many companies in this category (8 companies) leave the choice to employees and facilitate onsite working, often upon authorisation or desk booking, subject to limited availability. Further, four companies (C1, C7, C66 and C68) moved from employee choice to differentiated contracts with employees, including a clause that customers may request in-office working in selected projects, or, as in the case of Brazilian DB (C7), requesting those employees living nearby to visit the office regularly.

The other group comprises remote-only companies with no offices and typically no opportunities for onsite working (see the bottom six companies in Fig. 6). In our sample, we have two small, one medium-sized, one large, and two extra-large companies with fully remote operation. Those with offices (two out of six companies) use them primarily for administrative purposes, while three companies (C6, Stone Co (C6), and Webdraw (C9) in Brazil) can accommodate onsite work upon request. Out of these six remote companies, three have been designed to be remote. Keil KTM (C12) in Germany was fully remote before the pandemic. Stone Co (C6) in Brazil believes in the hyper-productivity of remote working and has institutionalized remote working up to the higher management levels. Webdraw (C9) in Brazil is a recently established company designed to be remote. Swedish factor10 (C53) and Brazilian C62 and C64 became fully remote during the pandemic, and as a result of successful forced work from home decided to close their offices and rent space for company events when needed. Finally, all companies in this category employ people living at a distance. One key reason for being remote is the explicit strategy to engage people from anywhere, even globally (in four out of six companies).

## 4.4. Companies with no clear orientation

In Fig. 7, we summarize how office presence is regulated in companies with no clear orientation towards either office or remote work. Interestingly, no location or company size predicts the companies in this group. The regulation of office and remote work in companies with no orientation also differs.

To our surprise, like in the remote-first companies, we found two companies with mandatory office presence, and one that set a recommended minimum for office days. The Italian companies (C20 and C27) in this category follow the national regulations promoting a minimum of half-time of onsite work, while Brazilian C63 introduced two office days to ensure engagement.

Three companies in the group with no clear orientation have decentralized regulation with leaders, departments, teams, or customers mandated to decide whether and when employees are expected to work onsite.

Finally, five companies offer employees the freedom of choice for work location. This freedom can mean full flexibility daily or be limited by certain conditions. An interesting company in this respect is multinational C15, where employees have to choose between office-first and remote-first mode, which becomes a part of the formal agreement for a 12-month period. It is a practical tool to keep track of the employees' needs for an office desk and maintain the office space of a reasonable size — not too large so that it is empty and not too small that employees who prefer to work onsite cannot find a desk.

The challenge with the office space in different companies is dealt with differently. An alternative strategy, for example, is introduced in German C11, which tracks the usage of offices for one year to decide whether to keep them as a coworking location. New offices in C11 are only opened if at least six employees are responsible for basic office management tasks.

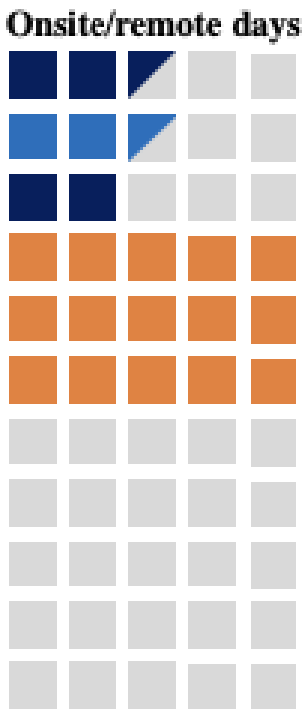

| Case | Size | Location | Required office presence | Fully remote positions | Recruitment |
|---|---|---|---|---|---|
| C20 | L | Italy | Half of the total working days of the month + 1 (10-11 onsite days/month) | As an exception | Within the country |
| C27 | S | Italy | 2-3 days/week of office work are recommended | No | Within commute distance |
| C63 | XL | Brazil | 2 days/per week onsite, additionally based on business necessity | Yes | Within the country |
| C48 | XL | Sweden | Work location decided with the manager, the team and the customers | Yes | Within the country |
| C33 | XL | Norway | Departments, projects and teams are mandated to decide | No | Within the country |
| C10 | M | Denmark, Sweden | No centralized requirement leaders are mandated to require office presence | Yes | Internationally |
| C15 | XL | International | Work from anywhere | Yes | Internationally |
| C11 | L | Germany | Employee choice | Yes | Within selected countries |
| C56 | M | US | Employee choice for seniors; juniors are expected to work 4 days/week onsite | Yes | Internationally |
| C13 | M | Germany, US | Employee choice | Yes | Within the country |
| C25 | S | Italy | Employee choice | Yes | Within the country |

**Fig. 7:** Work regulation in companies with no preference for office and remote working.
Each row represents a company's in-office work requirements across the workdays of a typical week (Mon–Fri), illustrated through color-coded blocks: Mandated office days, Recommended office days, Flexible days (employee-choice), and Days with decentralized office presence, subject to department, project, or team decisions. In some cases, specific weekdays are enforced, while in others, only the number of required in-office days per week is prescribed.

Accommodating a high degree of freedom in choosing between on-site and remote work comes, in principle, with challenges when designing the office spaces. Several companies reported attempts to redesign their offices to separate silent areas for focused work, and dedicated areas for socialization and team collaboration. The high fluctuation in the number of employees using the offices (in C11, this could be anything between 5-7 on average to over 60 in peak times) rendered the office redesign endeavour to accommodate the flexibility cumbersome.

## 4.5. Motivations for regulation

To understand why companies have chosen a particular work regulation, we summarized the top motivations given by the interviewees. Explaining their work regulation, companies could give multiple reasons for encouraging or mandating office days, or providing work flexibility and remote work opportunities. Fig. 8 shows the resulting categories and how our analysis of the reasons for office and remote work reveals a clear divergence in motivational drivers (see Fig. 8).

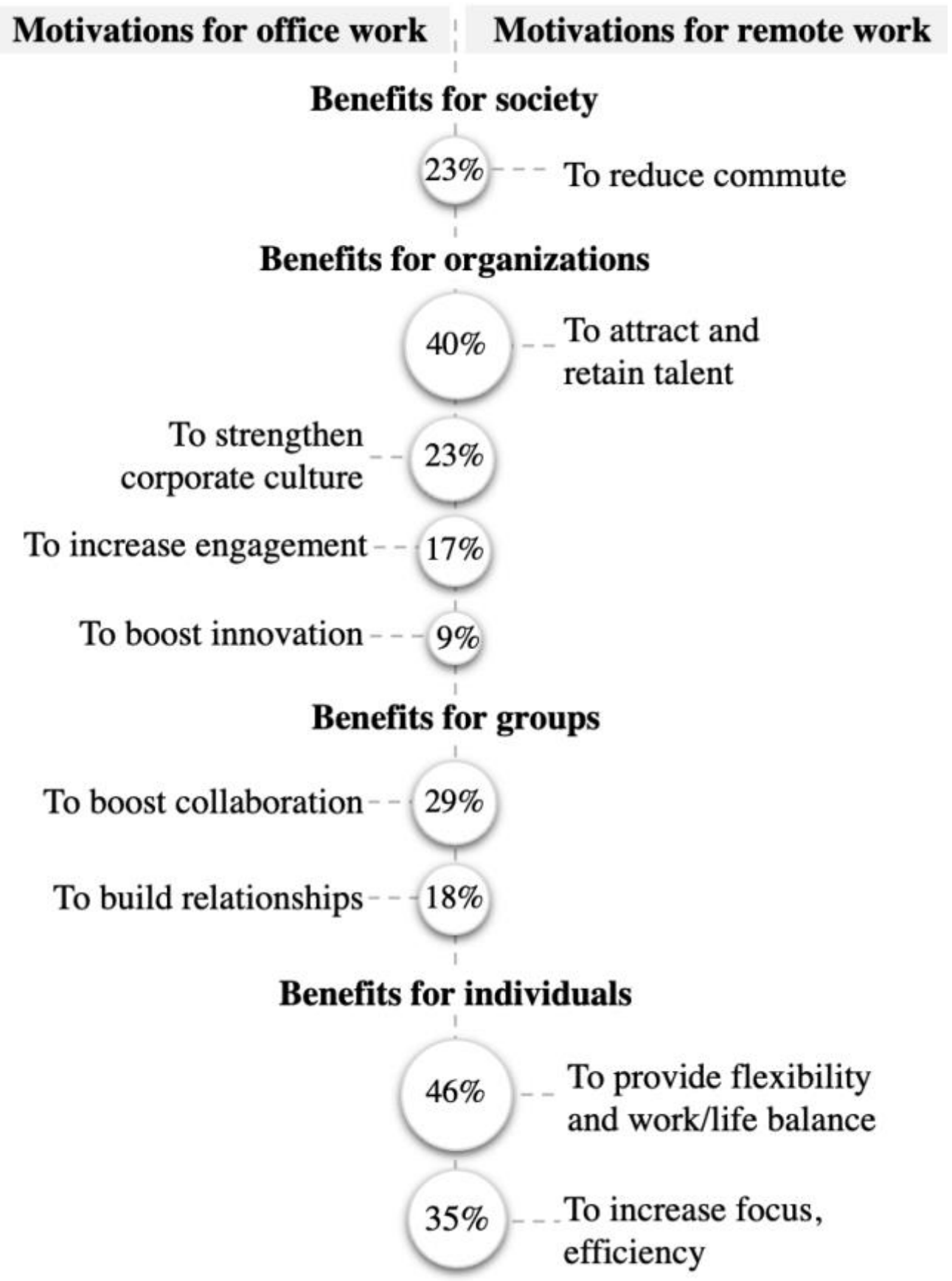

**Fig. 8:** Top motivations for regulating office days and for allowing remote work, with the percentages representing the share of companies in the whole sample

The top motivations for encouraging or mandating office work are largely rooted in factors that benefit the organization (corporate culture, engagement, innovation) and work groups (collaboration and social relationships). In one instance, office presence was seen as the place to strengthen the culture, solve problems efficiently, and innovate together. Co-location is expected, in particular, to boost collaboration, mutual help, and build empathy and deeper interpersonal connections.

In contrast, motivations for remote work tend to center on individual-level benefits (work-life balance, focus, and efficiency). However, companies oriented toward remote work also mentioned talent attraction and retention, and contribution to sustainability by reducing commute, which together show that remote working also supports broader organizational and societal goals.

Most companies in our sample accommodate hybrid working, a combination of office and remote work, and mention reasons on both sides of the spectrum. Some of these companies specifically mention flexibility—offering employees the possibility to decide whether to work onsite or remotely as one of the important reasons for being hybrid.

## 4.6. Changes in remote work regulation

Our analysis indicates that most companies have not changed their remote work policies established after the pandemic. In our sample, 19 companies (28%) have made adjustments in response to evolving organizational needs by introducing new conditions or exceptions in regulating either onsite or remote working, or increasing required office presence in the policy (upward trends in Fig. 9).

We did not observe companies moving toward increased remote work flexibility (no downward trends in Fig. 9) except for the Italian Klopotek (C24), which has made a series of changes with the largest upward jump from employee choice to recommending three office days in November 2024, and then landing on a two office day policy in January 2025. The strict policy was received with significant pushback, and after monitoring the reaction of the employees for one month, the company agreed to lower the return-to-office mandate requirements.

The twelve companies (approximately 18%, marked in Fig. 9 in yellow with upward trajectories) that have shifted their policies toward requiring additional in-office days compared to their initial post-pandemic policies span different sizes, industries, and geographic locations. In some cases, these changes were motivated by making the office policies more concrete. For example, C18, C44, and C49 moved from the rather vague 50% yearly presence to a stricter and more concrete 60% weekly office presence. Interestingly, most changes landed on three in-office days/week, making it the most prominent category in our sample. The main motivation among these companies was to increase co-presence and the amount of face-to-face interaction. Similar motivation concerned the full return to the office in three companies with significant upward jumps: Brazilian Maxtrack (C5), Italian C59, and Indian C67. Interestingly, like Klopoteck (C24), C67 also reported considerable pushback from the employees, followed by voluntary resignations and difficulties in hiring new employees. However, in all three cases, the new policy remains.

Shifting to more office presence in practice was not always an easy task. For example, Indian C49 moved to three in-office days/week after downsizing their office space. As a result, the company introduced a transition period with two office days plus three remote days as a temporary measure until additional office space becomes available. Office space limitations pushed the Indian C16 to restrict further the flexibility in choosing office and remote days, as they had to plan the teams to work in shifts. Italian C21 moved from three office days to requiring employees to include Monday or Friday as one of the office days to balance office occupancy throughout the week. This one and other companies that changed their policies, even though not moving along the Y-axis, are also marked in yellow in Fig. 9, to indicate the proportion of companies with changes (26%) compared to those that have not changed their work regulation (74%).

Five companies that have introduced changes shifted toward strategic reframing of policy language. For instance, the Swedish C46 evolved from a vague requirement of "majority of the time in the office" to a more specific interpretation—"three days a week in the office" in spring 2024. Similarly, the Swedish C47, which in its initial post-pandemic policy centered on a "3+2 principle" (three office days plus two remote days), removed the explicit emphasis on the two remote days and communicated the requirement of "the majority of time onsite."

Further, a notable trend is the introduction of differentiated policies for different employee categories or specific conditions, particularly for new hires and those between client projects, or in projects where clients require office presence. For example, multinational C15 across locations has implemented a tiered approach where junior-level new hires are expected to work onsite during the first year of employment, while for more experienced new hires, this period is only a month long. Notably, such an approach also exists in other companies that have not changed their regulation but foresaw the need for onsite onboarding, like C56 in the US, where juniors must visit the office four days/week. Another example of policy adjustments concerns employees without active customer projects or between assignments, who are now expected to work primarily onsite (the Norwegian consulting companies Knowit (C37) and Kantega (C38)).

Beyond daily or weekly presence requirements, some companies have introduced synchronized periods to ensure regular in-person collaboration across the organization. For example, C15 implemented "core weeks" (once or twice per year), during which in-office presence is mandatory even for employees living remotely from the offices. Their approach is expected to provide quality time for collaboration while maintaining a high degree of flexibility throughout the rest of the year.

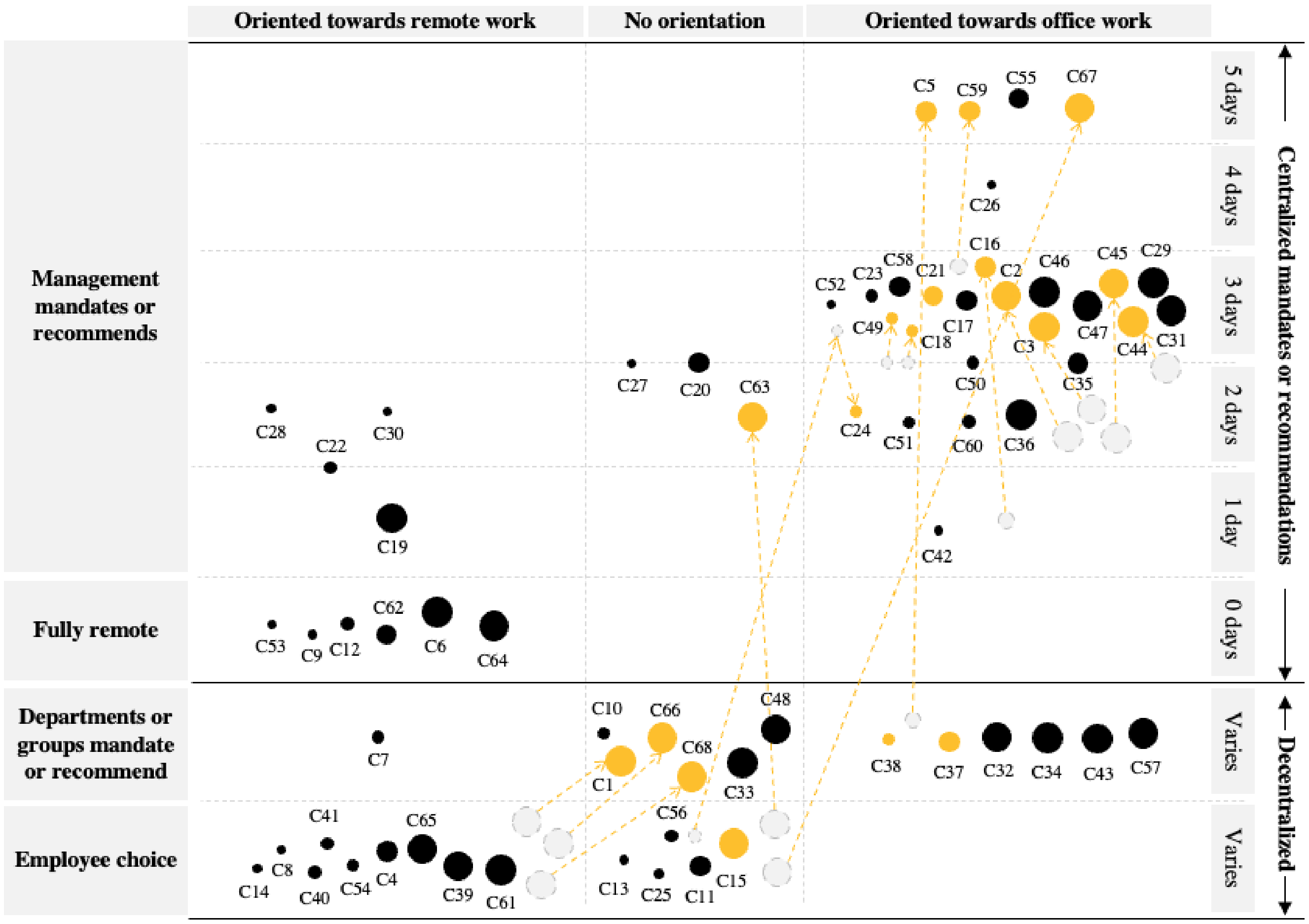

**Fig. 9:** Work regulation in remote-first companies. Companies are visualized based on size: • Small • Medium ● Large ● Extra large. The changes are highlighted with dotted lines between grey bubbles (the initial state) and yellow bubbles (the current state of regulation; also used for companies that have introduced changes, but did not change their location in the graph),

The state of remote work regulation in Fig. 9 is not final. We received indications that several companies are working on adjustments to their policies. For example, Scandinavian C10 is preparing a new policy that may include stricter office presence requirements. Brazilian Electrobras (C3) has moved from two to three office days a week during our study, and has already suffered from the negative consequences, loss of talented employees who were unwilling to adapt to the new strategy. The changes in many other companies are delayed due to the fears of a negative reaction and potential resignations. It is true for the Italian C19 that temporarily put the shift to full-time office presence for junior new hires on hold. Another reason against the differentiated policy in their case is the fact that mandatory office presence for juniors may be less meaningful without requesting the presence of senior mentors.

Overall, most of the reported changes are incremental adjustments rather than radical shifts, with a few mentioned exceptions. Companies experiment, learn, and adjust their policies based on specific operational needs and physical constraints without changing their fundamental orientation. Notably, most of the changes appeared in companies oriented towards office work or in companies that had no orientation and, based on their experiences, shifted towards an office-centered culture. Only a few remote-first consultancies had to change their principles and introduce the possibility of in-office work based on customer requests, landing on our scale in the "No orientation" position.

## 4.7. Recruitment and personnel management

The changes in remote and onsite work regulations following the pandemic brought new considerations to the forefront, particularly regarding recruitment strategies and employee relocation. Whereas, in the past, individuals typically lived near their workplace, today employees are not necessarily obliged to resign when moving to a different city or even when changing countries. It is largely due to the growing acceptance of fully remote work arrangements. In Fig. 10, we illustrate how the case companies with different work orientation models, from office-first to remote-first, approach recruitment and whether they offer fully remote positions.

Our data shows that companies oriented towards office work predominantly recruit candidates who live within commuting distance or in the same country. Fully remote positions in such companies are either not offered at all, or provided as rare exceptions, something negotiable, reserved for employees with special needs or unique circumstances, rather than the norm.

In contrast, companies oriented towards remote work will likely offer fully remote positions as a standard practice. Although most remote work-oriented companies recruit from within the country, it's not uncommon for them to also span national boundaries, reflecting a broader and more flexible talent acquisition approach.

Recruitment strategies in companies with no clear orientation in our sample vary but are more favourable to fully remote working than those oriented towards the office and are also open to recruitment beyond national borders.

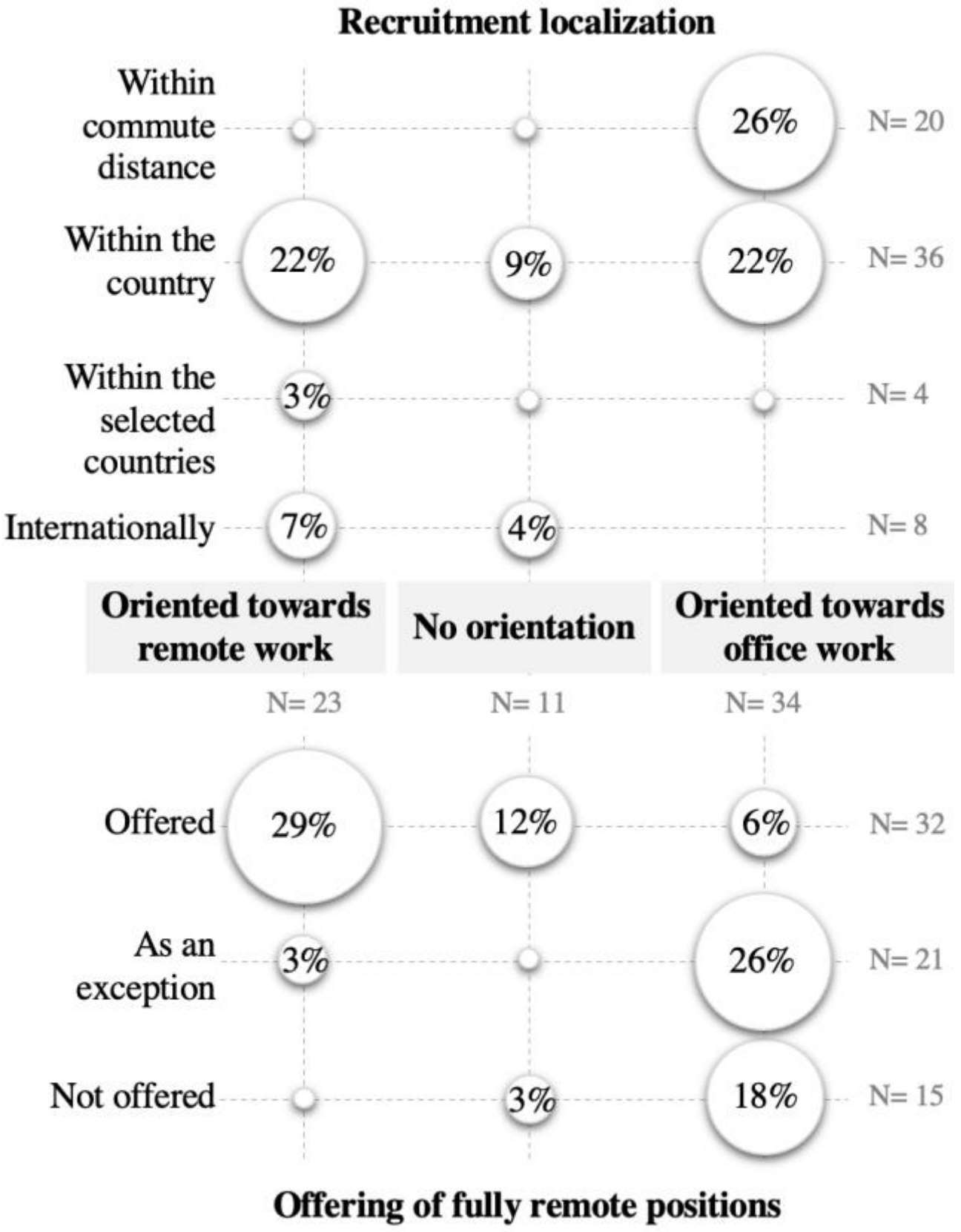

**Fig. 10:** Summary of the prevalence of fully remote positions in companies by orientation towards remote vs office work

Notably, two primary drivers behind companies embracing remote work and expanding their recruitment efforts beyond the boundaries of traditional commute distance include access to a broader talent pool and increasing corporate attractiveness as an employer. By offering remote positions, companies can tap into skilled professionals regardless of location and make the corporate image more competitive in the eyes of top-tier candidates. These are 8 (11%) of the companies in our sample. Additionally, companies have an economic incentive to adopt broader recruitment strategies, as companies can benchmark employee salaries against local IT job markets. Nine (13%) companies, including those leveraging diversity of economic conditions within the national borders, adjust their salary setting based on employee location. We foresee that exploiting wage differences may increase the interest in offering fully remote positions as a common means for labor cost reduction, and vice versa. However, the vast majority of companies in our sample, at the moment of our investigation, do not report any differences in salary setting or explicitly adopt an equal pay strategy, offering standardized compensation based on career performance and disregarding the employee's location. The latter approach exemplifies companies that promote fairness and reinforce a consistent organizational culture.

# 5. Discussion

## 5.1. Hybrid work is the "new norm"

Our findings reveal that the heavily publicized full-time RTO mandates seen in major US tech companies are not a significant trend in our sample. While companies like Amazon and X have implemented full-time return-to-the-office policies, our broader sample, considering company size and locations, reveals a more nuanced reality. Yet, we can conclude that there is a shift towards increased office presence in the past four years, as also reported by others (Ding and Ma, 2023; Eng et al., 2024). ⅓ of the companies in our sample adopted hybrid work policies like Apple, Microsoft, and Meta, requiring office attendance of 50-60%. However, in contrast to studies that advocate for structured management practices and fixed onsite work schedules to maintain innovation, productivity, and organizational cohesion (Eng et al., 2025), our findings show that RTO in many of the cases we studied is framed as a preference rather than enforced as a uniform, mandatory policy. Some companies in our sample explicitly step away from having a unified RTO mandate that are typically reported to cause employee dissatisfaction (Eng et al., 2024), and instead opt for decentralized policies. More importantly, our findings suggest that companies may have an orientation towards office work but permit arrangements with highly flexible conditions (see regulation with few office days or issuing recommendations for the expected amount of office presence in Fig. 4). These companies follow what is recognized as a trust-based strategy that communicates managerial expectations (Gajendran and Harrison, 2007) with a clear emphasis on in-person collaborative behaviors rather than a proportion of days with and/or without appearances in the office.

Before the pandemic, "the norm" was office-oriented work with tech companies allowing some flexibility and remote-first companies being mere outliers. Today, the new norm is hybrid working. In our sample, 85% of the companies practice hybrid working to different degrees, with the remaining 15% divided between remote-only (9%) and office-only (6%) companies. Hybrid, indeed, is here to stay (Allen et al., 2024).

Admittedly, there is no clear "best practice" regarding the right degree of office vs remote working. Companies are left to find their ideal model within their possibilities, and legislators are left to implement certain foundations for protecting the employees, guiding principles for such areas as insurance and taxation, and ensuring a healthy population. Even after three-four years of experimentation, there is significant variation in how companies regulate where work is expected to be done. After the pandemic, some countries have returned to pre-COVID remote work regulations, formally consisting of an individual agreement between the company and the employee, integrated within the general labour contract. Others have evolved or established remote work regulations based on the experience acquired during the COVID pandemic and because of expanding the recruitment strategies.

## 5.2. Negotiating the expectations

Reflecting on the companies that changed their policies, our findings reveal a notable unidirectional pattern. They consistently moved toward increased office presence when changes occurred rather than expanding remote work flexibility. This trajectory mirrors trends observed in high-profile tech companies, as illustrated in Fig. 1, where organizations like Apple, Microsoft, and Meta have implemented stricter office attendance policies, and confirms prior research suggesting that companies, especially large ones, adopt structured hybrid policies and increase the requested office presence (Ding and Ma, 2023; Eng et al., 2024). The popularity of the three office days per week as the most common RTO mandate in our sample aligns with the findings of Ding and Ma (2023). Interestingly, while right after the pandemic 15% of companies in our sample introduced three office days, by 2025 26% of companies in our sample have tried out this work arrangement.

Related research found that RTO mandates are a common go-to solution to the managerial challenges and stress caused by the attempt to satisfy employees' different preferences for collaborative and social interactions (Eng et al., 2024). Yet, when employers change the corporate work regulation, the employees' readjustments to the new rules do not always happen at ease. The pandemic has fundamentally altered employees' expectations, giving rise to new psychological contracts — the unwritten, perceived reciprocal obligations between employees and employers (Rousseau, 1989). Employees with a higher work-from-home preference value the ability to work remotely and likely expect that the company will provide the resources necessary to institutionalize this work modality, leading them to expect more relational obligations from their employers (Gutworth et al., 2024). As such, remote work has converted from an exclusive privilege that managers granted to the selected few (Olson, 1983), to a baseline expectation for many (Smite et al., 2023), at least in tech companies where developers perceive being able to work from anywhere. Mandating a return after employees have adapted to remote work can be perceived as a breach of the psychological contracts (Rousseau, 1989), resulting in intense emotional reactions. It explains why return-to-office mandates often result in decreased job satisfaction (Ding & Ma, 2023) and increased turnover or threats of resignation (Allen et al., 2024; Barrero et al., 2021a; Eng et al., 2024; Van Dijcke et al., 2024), particularly among high-performing employees (Elliott, 2024).

Our findings confirm that reversed flexible policies and enforced structured in-office attendance is met with strong employee resistance. For example, the Italian Klopotek moved from flexible, employee-driven decisions to three mandatory office days—only to backpedal to two in-office days after employee pushback. Such negative responses, including threats to resign, mirror patterns identified in prior research (Allen et al., 2024; Barrero et al., 2021a).

Job candidates today are attuned to how companies regulate remote work. Human resources professionals report that remote work policy is often the first topic raised during job interviews, especially among top talent. Fully on-site companies risk appearing less attractive to candidates unless they can clearly justify their policies. In our sample, even companies that identify as office-first, offer fully remote roles—at least on an exceptional basis.

These findings demonstrate that companies that prefer in-person interactions today face a complex trade-off. On one hand, structured office attendance may help maintain their collaborative culture and ease managerial burden, on the other hand, it can reduce their attractiveness to current and prospective talent. With flexibility being perceived as a core part of the employment contract, companies introducing RTO initiatives must develop convincing arguments to support them.

For many employers, structured hybrid arrangements, particularly with three in-office days per week, serve as a compromise solution that resolves the conflict. Each side gets something. Notably, while less structured approaches or few office days required result in fragmented team presence (Moe et al., 2023), the three-day policies offer predictability, help justify office infrastructure, and give managers a clear mandate to rely on without needing to negotiate it on an individual basis (Tabahriti, 2022). Even though such policies may not fully satisfy all employees, they become a "fair-for-all" fallback that minimize frictions and offer a baseline that can be fine-tuned in exceptional cases.

## 5.3. Beyond one-size-fits-all: Differentiated policies

Following the tension between strict RTO mandates and negotiated flexibility, a key finding in our study is the rise of differentiated work policies. Although the state of work regulation in Fig. 4 highlights the popularity of the three-day return-to-office mandates (24%), 19% of companies, regardless of corporate orientation, allow departments, teams, or projects to determine where to do the work. In several cases, work arrangements are dynamic, adjusted on demand or situational basis. Further, 7% provide a recommendation for office presence, a target that can yet accommodate individual circumstances. Finally, 21% of companies let employees decide where to work. These companies (46% in total) explicitly state they do not subscribe to a one-size-fits-all solution and instead prioritize contextual decision-making.

Even companies with centralized policies do not always apply the same rules uniformly. Some have conditional guidelines that take into account employee roles, seniority, workload, or proximity to the office. The rationale behind these choices aligns with the organizational control theory (Ouchi, 1979). According to Ouchi (1979), decision rights should ideally reside where the relevant information is most accurate and immediate. For companies like C32, C33, Storebrand (C34), C48, and C57, decentralizing remote work decisions to the team level acknowledges that teams are best positioned to assess their collaboration needs. Differentiated approaches in the office-first consulting firms like the Norwegian Knowit (C37) and Kantega (C38), and also remote-first C1, C66, and C68, prescribes adjustments based on client expectations and/or project status, exemplifying the contingency-based reasoning that organizations should adapt their policies to fit specific operational demands (Lawrence & Lorsch, 1967). Similar approach is evident in C15 and C56, who have tiered remote work conditions for new hires, illustrating strategic workforce differentiation (Huselid & Becker, 2011). Delegation of locational decisions to employees, in its turn, reflects the principle of enabling formalization (Adler & Borys, 1996), where flexible organizational rules and policies are designed to empower employees rather than constrain them.

In other words, companies with diverse business functions and/or workforce realize that uniform, one-size-fits-all policies are often inadequate for addressing the distinct collaboration needs of different projects, teams, or roles.

## 5.4. Recommendation for deciding where work should be done

Our analysis of the current state of work regulation did not reveal any clear "winning" policy. It was not our goal either. Rather than advocating for a universal solution, our research aims to paint the diverse, nuanced and evolving landscape of modern work policies. We identified a broad spectrum of emerging practices shaped by organizational goals, employee expectations, and physical, cultural or legal constraints. It is also worth emphasizing that company size does not seem to be related to company policy. Based on this nuanced understanding, we offer the following advice for companies that are establishing or reassessing their work policy.

For companies that consider the office as the primary location of work, we emphasize that **orientation towards office work does not require strict mandates**. Instead of strict mandates, companies may introduce a trust-based approach that puts a **recommended target** for particular office days or office presence in general. If you believe that mandates are irreplaceable in your context, you can further question whether it should be a one-size-fits-all mandate or a **differentiated approach** based on contingency reasoning that recognizes the diverse needs for different roles, seniority, or project circumstances. Differentiated approaches include exceptions from the main policy or diverse set of rules.

For companies that have or consider downsizing the office so it can only fit a fraction of employees, we recommend **evaluating the consequences of shifting to predominantly remote work before reaching the point of no (or hard) return.** Consider whether your teams can maintain collaboration, innovation, and a shared sense of belonging without regular physical interaction. Ask: What might be lost if a shared space no longer exists, even part-time? Could occasional in-person gatherings or satellite hubs preserve culture and team cohesion in a distributed model? Several studied companies that initially had a high degree of flexibility reconsidered their work modality but were constrained by the decisions made about the office space, or expanded hiring strategies. To address the office space shortage, we identified the use of co-working spaces as a possible fall back solution. However, once a company starts hiring people on a large distance, reverting the orientation from remote working to required in-office presence might be hard. In our research, office-first companies kept the option of having fully remote positions as an exception.

Further, some organizations may deliberately **choose not to adopt a remote-first or office-first orientation**—and that can be a legitimate strategy in itself. Instead of defining the workplace at the corporate level, they delegate decision-making to business units, departments, projects, teams, or individuals, catering for the needs of those willing to work onsite and at the same time those willing to work remotely. This **decentralized, no-orientation strategy** allows for a more agile and responsive model, especially in contexts where business units serve different markets. In such models, alignment is achieved through shared principles and trust in decentralized decision-making.

In developing a policy, we also advise organizations to **explicitly consider the benefits they wish to prioritize**, as illustrated in our analysis (see Fig. 8). Different work models support different strategic goals. For example, flexible policies may enhance well-being and talent attraction, while in-person work may strengthen collaboration and corporate culture. By identifying and aligning with the most important outcomes for both the organization and its people, companies can avoid superficial copy-paste strategies and instead craft purpose-driven arrangements.

Furthermore, **organizations should recognize that policies are rarely absolute**. Even those with defined policies often make exceptions—for example, when hiring or when employees move. As such, companies should design policies with **built-in adaptability**, acknowledging that needs evolve over time and across roles. Moreover, framing a policy as “subject to periodic review” can help manage expectations and reduce resistance when updates are needed.

Finally, our main recommendation to all companies is to **experiment and adjust their policies and approaches** based on the output they provide and the feedback they receive. Companies that remain responsive are likely to be better positioned for long-term success.

## 6. Conclusions and future work

In this paper, we presented our findings from an analysis of post-pandemic work regulation in 68 companies varying in size, domain, and geographic location. Our results reveal a highly diverse landscape of work regulation, characterized by differences in orientation (office vs. remote), formality (formal policies vs. informal agreements), flexibility (the extent of employee choice), and control (centralized vs. decentralized decision-making).

Our results indicate that the majority of companies support hybrid work arrangements, and only four have returned to full-time office presence, and six have moved to full-time remote work, four of which have abandoned physical offices. Like prior research (Ding and Ma, 2023), we too found that larger firms are more inclined to adopt RTO policies. Our study reveals, however, that neither corporate size nor general orientation, whether office-first or remote-first, does not always predict actual work regulation. Many office-first companies do not enforce strict RTO mandates, while some remote-first organizations still maintain structured expectations around in-office presence.

We also conclude that many companies are still in a state of experimentation and adaptation. Since the initial announcements of post-pandemic work practices, several companies have revised the number of office or remote days or adjusted the conditions for onsite and remote work. Although no companies in our sample have formally increased flexibility since the initial mandates (with only one example of a company loosening the strict office attendance policy), many exhibit flexibility in applying work regulations. It includes decentralized policies and trust-based agreements allowing recommended or fully autonomous decisions regarding work location.

Future work should focus on the consequences and effectiveness of the diverse regulatory approaches. One promising direction is to assess the actual outcomes of the various regulations and validate whether the corporate motivations are justified. Particularly, we suggest evaluating strict return-to-office mandates with clearly defined in-office days, whether these policies effectively increase office presence, and at what cost. Comparative studies could examine whether trust-based, flexible policies, especially those complemented by workplace redesign and additional perks, yield equal or superior outcomes regarding employee engagement, collaboration, and performance. Equally important is understanding the long-term impact of strict mandates on employee satisfaction and retention. Moreover, future studies should move beyond binary classifications of policies and explore the nuanced features of work regulation, as found in our study. For example, return-to-office mandates may differ in the number of required office days and how the locational decisions are made. Evidence from evaluative studies on various work regulation approaches is especially valuable in the current landscape, where established theories and practical guidelines are lacking due to the unprecedented scale and speed of post-pandemic workplace changes.

## Acknowledgements

We would like to thank the companies and our study interviewees for participating in our research and for the sincere interest in advancing the common understanding of work regulation trends.

## Author contribution declaration

Darja Smite and Nils Brede Moe *designed the study*. All authors were involved in *data collection* and *interactions with case companies*. *Data analysis* was performed by Darja Smite and Nils Brede Moe, with the help of Guilherme Horta Travassos and Teresa Baldassarre. *Data visualisations* were designed by Darja Smite. The manuscript was *drafted* by Darja Smite, Nils Brede Moe, and Fabio Calefato, and all authors have provided *critical feedback* and *final approval* of the manuscript.

## Declaration of interests

**Funding:** This study was funded by the Swedish Knowledge Foundation through the projects S.E.R.T. (grant number 2018/010), WorkFlex (grant 2022/0047), and Research Council of Norway through the Transformit project (grant number 321477). Prof. Travassos is supported by the Brazilian grants from CNPq and CNE FAPERJ.

**Conflict of interests:** The authors have no competing interests to declare.

## Data availability statement

The policies collected during the study and the interview data are not available for confidentiality reasons.

## Declaration of generative AI and AI-assisted technologies in the writing process

During the preparation of this work the authors used ChatGPT 4.0. AI assistance was limited to language editing, paraphrasing and grammar correction. AI assistance was strictly supervised, the generated content was reviewed and edited as needed. All critical analysis, arguments, and conclusions presented in this paper are the result of the authors' analytical work. The authors take full responsibility for the content of the paper.